\documentclass[prb,twocolumn,preprintnumbers,amsmath,amssymb,floatfix,longbibliography,nofootinbib,superscriptaddress]{revtex4-1}
\usepackage{graphicx}
\usepackage{graphics}
\usepackage{psfrag}
\usepackage{hyperref}
\usepackage{multirow}
\usepackage[sort&compress]{natbib}
\usepackage{amssymb}
\usepackage{amsmath}
\usepackage{amsthm}
\usepackage{bbold}
\usepackage{bbm}
\usepackage{enumerate}

\newcommand{\rmi}{{\rm i}}

\begin{document}

\hypersetup{pdftitle={Ground state of the three-dimensional BCS d-wave superconductor}}
\title{Ground state of the three-dimensional BCS d-wave superconductor}

\author{Igor F. Herbut}
\affiliation{ Department of Physics, Simon Fraser University, Burnaby, British Columbia, Canada V5A 1S6 }
\author{Igor Boettcher}
\affiliation{Joint Quantum Institute, University of Maryland, College Park, MD 20742, USA }
\author{Subrata Mandal}
\affiliation{ Department of Physics, Simon Fraser University, Burnaby, British Columbia, Canada V5A 1S6 }

\begin{abstract}
We determine the mean-field ground state of the three-dimensional rotationally symmetric d-wave ($\ell=2$) superconductor at weak coupling. It is a non-inert state, invariant under the symmetry $C_{2}$ only, which breaks time reversal symmetry almost maximally, and features a high, but again less-than-maximal average magnetization. The state obtained by minimization of the expanded sixth-order Ginzburg--Landau free energy is found to be an excellent approximation to the true ground state. The coupling to a parasitic s-wave component has only a minuscule quantitative and no qualitative effect on the ground state.
\end{abstract}
\maketitle

\section{Introduction}

The problem of Cooper pairing with higher angular momentum and the concomitant superconducting state arises often in many-body physics, with the p-wave state ($\ell=1$) in $^3$He probably being the best known example.\cite{vollhardt} When pairing occurs in the spin-singlet channel and the usually dominant s-wave state ($\ell=0$) is suppressed due to electron-electron interactions, pairing in the d-wave ($\ell=2$) channel ensues. The multi-component complex order parameter that describes situations with $\ell>0$ typically leads to the problem of finding the optimal configuration which minimizes the free energy within a large manifold of possible realizations. The expectation is, however, that the state that minimizes the energy still displays some residual symmetry.\cite{bruder} Since the set of continuous and discrete subgroups of the original symmetry group SO(3) in three dimensions is limited, this guiding principle greatly simplifies the search for the ground state. Identifying the ordered ground state in the case of multiple complex field components is also relevant for spinor Bose--Einstein condensates, where the degrees of freedom are bosonic atoms.\cite{spinorBEC}

A recent candidate for an $\ell=2$ superconductor is the half-Heusler compound YPtBi, where the temperature dependence of the penetration depth indicates unconventional pairing, and the Fermi level is close to the quadratic band touching point.\cite{brydon,paglione} If  the Fermi level would be precisely at such a ``Luttinger point" of the band structure \cite{luttinger, moon, janssen1, janssen2, boettcherSC}, than the superconducting d-wave state predicted from weak-coupling theory would preserve time reversal symmetry, and the ground state would be the uniaxial nematic state with line nodes in the spectrum.\cite{boettcher} Away from such a point, as it is typically the case in real materials with nonzero carrier density, the Ginzburg--Landau free energy derived at weak coupling suggests that the d-wave ground state breaks time reversal symmetry maximally, but at least at the quartic level leaves the question of the actual ground state open.\cite{brydon} This is due to the accidental vanishing of one of the three symmetry-allowed quartic terms that would otherwise break the degeneracy between the  time-reversal-symmetry-breaking states.\cite{mermin} In this situation, at least within the confines of the weak-coupling and  Bardeen--Cooper--Schrieffer (BCS) mean-field theory, one is forced to consider the next-order, sextic terms in the Ginzburg--Landau expansion in terms of the d-wave order parameters, and search for the minimum of the free energy within a rather large manifold of states.\cite{boettcher} Further pairing mechanisms for YPtBi that may arise from the Fermi level being away from the band touching point have been classified and compared in Refs. \cite{meinert,agterberg,wu,savary,venderbos,mandal,liu,roy,kim,szabo}.(See also \cite{barzykin, mazin}, for earlier related work.)

In this paper we consider the general problem of spontaneous breaking of the rotational SO(3)  and particle number U(1) symmetries by an $\ell=2$ superconducting state at weak coupling and at a finite chemical potential, when it suffices to consider the Hamiltonian projected onto the Kramers-degenerate low-energy band at the Fermi level. Going beyond the usual Ginzburg--Landau expansion, and minimizing with respect to the norm of the Cooper pair wave function first, we find that the BCS ground state at $T=0$ ultimately maximizes a specific integral over the Fermi surface of the Cooper pair internal wave function. Utilizing Michel's theorem \cite{michel} in the search for the global minimum of the energy, we find the ground state to be invariant under the smallest subgroup of the rotational group, namely $C_2\simeq \mathbb{Z}_2$. Minimizing within the parameter space of $C_2$-symmetric states we find that the exact Cooper pair ground state a) is {\it nearly} orthogonal to its time-reversed copy, i. e. breaks time reversal symmetry, but not quite maximally, and b) exhibits a large, but again less than maximal expectation value of the orbital angular momentum $\langle \textbf{L} \rangle^2$, and therefore of the magnetization.

Checking against the usual Ginzburg-Landau expansion at finite temperature we find that an excellent approximation to the exact ground state is selected by the sextic term in the free energy.\cite{boettcher}  Essentially the same superconducting state is therefore preferred at {\it all}  temperatures below the critical temperature. We also show why including the symmetry-allowed coupling to the parasitic s-wave component \cite{kim} in principle modifies the ground state quantitatively, but only minutely so and not at all qualitatively. The specific features of the low-temperature superconducting state such as magnetization and time-reversal symmetry breaking can be accessed in experiment, for instance, through
magneto-electric effects \cite{smidman}, surface excitation spectra \cite{menke}, or optical conductivity \cite{boettcherOPT}.

Our analysis is organized as follows. We first introduce the SO(3)-invariant BCS model for a parabolic band with pairing occurring in the d-wave channel, together with several representations for the five-component complex order parameter. We then derive an exact functional in the weak-coupling limit that determines the ground state of the model at zero temperature and establish the solution to the corresponding optimization problem with the help of classes of states that transform under the subgroups of SO(3). We compare the ground state to the result of optimizing the Ginzburg--Landau free energy expanded to sixth order at low temperatures. Eventually we compute the quantitative effect of a parasitic s-wave component on the ground state.

\section{$l=2$ pairing and the order parameter}

Let us begin with the Lagrangian in standard three-dimensional BCS form with the pairing interaction between the time-reversed states in the spin-singlet channel given by
\begin{align}
 \nonumber &L(\tau) = \sum_{\sigma=\pm} \sum_{\textbf{k}} \Psi_{\sigma}^*(\tau,\textbf{k}) ( \partial_\tau + \xi_{\textbf{k}}) \Psi_{\sigma} (\tau,\textbf{k})\\
 \label{eq1} &{}- \sum_{\textbf{k},\textbf{p}}\hspace{-0.5mm}{}^\prime\ g(\textbf{k},\textbf{p} ) \Psi_{+} ^* (\tau,\textbf{k}) \Psi_{-}^* (\tau,-\textbf{k}) \Psi_- (\tau, \textbf{p}) \Psi_+ (\tau, -\textbf{p}),
\end{align}
where $\xi_{\textbf{k}} = k^2 /(2m) -\mu$. The pairing interaction is assumed to be attractive in the d-wave channel,
\begin{equation}
 \label{eq2} g(\textbf{k},\textbf{p}) = g P_2(\hat{\textbf{k}}\cdot\hat{\textbf{p}}),
\end{equation}
with $P_2 (x)=\frac{1}{2}(3x^2-1)$ the second Legendre polynomial and $g>0$. The prime on the second sum in Eq. (\ref{eq1}) as usual implies that only the momenta within a cutoff $\Lambda \ll k_{\rm F}= \sqrt{2m\mu} $ around the Fermi surface are to be included. $\Psi_\sigma(\tau, \textbf{k})$ are the usual Grassmann variables. The Lagrangian $L(\tau)$ represents the simplest rotationally-invariant BCS model for $\ell=2$ pairing. Complementary, it describes spin-orbit coupled materials with a four-band quadratic band touching point close to the Fermi level, described by the Luttinger Hamiltonian, with complex tensor order pairing between the electrons of total angular momentum 3/2,\cite{boettcher} {\it projected} onto the two Kramers-degenerate bands that cross the finite chemical potential.\cite{brydon, venderbos, roy}

Using the addition theorem for spherical harmonics,\cite{QM} Hubbard--Stratonovich decoupling of the interaction term, and applying the mean-field approximation to integrate out the fermions in the background of a constant order parameter,\cite{negele} the mean-field superconducting state is given by the minimum of the effective action
\begin{equation}
\label{eq3} S[\vec{\Delta}] =  \frac { |\vec{\Delta}|^2 }{g} - T \sum_{\omega_n,\textbf{k}}\hspace{-0.6mm}{}^\prime\   \ln \Bigl( \omega_n ^2 + \xi_{\textbf{k}}^2 + |\Delta_a d_a (\hat {\textbf{k}} ) |^2 \Bigr).
\end{equation}
Here $\vec{\Delta}=(\Delta_1,\Delta_2,\Delta_3,\Delta_4,\Delta_5)$ comprises five complex order parameters that transform under the $\ell=2$ representation of SO(3), $\omega_n = (2 n +1)\pi T$ are the Matsubara frequencies, and the five functions $d_a(\hat{\textbf{k}})$ are real spherical harmonics given by
 \begin{align}
 d_1 &= \frac{\sqrt{15}(k_x^2-k_y^2)}{2k^2},\ d_2 = \frac{\sqrt{5}(2k_z^2-k_x^2-k_y^2)}{2k^2},\\
 \label{eq4} d_3 &= \frac{\sqrt{15}k_zk_x}{k^2},\ d_4 = \frac{\sqrt{15}k_yk_z}{k^2},\ d_5 = \frac{\sqrt{15}k_xk_y}{k^2}.
\end{align}
We normalize the functions so that the angular average over the sphere defined from $|\hat{\textbf{k}}|^2=1$ yields $\int \frac{d\Omega}{4\pi} d_a d_b = \delta_{ab}$. We implicitly sum over repeated indices, and in our units $\hbar = k_{\rm B} =1$. The quasiparticle dispersion for excitations close to the Fermi level that results from Eq. (\ref{eq3}) is given by
\begin{align}
 E(\textbf{k}) = \sqrt{\xi_{\textbf{k}}^2+|\Delta_ad_a(\hat{\textbf{k}})|^2}.
\end{align}
Typically, the action $S[\vec{\Delta}]$ is expanded in a Taylor series in powers of $\Delta_a$, which, when truncated at certain order, leads to the usual symmetry-dictated Ginzburg--Landau expression. At $T=0$, however, one can actually dispose of the expansion. To this end, we first introduce some helpful notation for the representation of the order parameter.

Every order parameter $\vec{\Delta}$ can be understood as a state $|\vec{\Delta}\rangle = \Delta_a |M_a\rangle$ in a five-dimensional Hilbert space, where the $|M_a\rangle$ constitute the $\ell=2$ real basis, satisfying $\langle \hat{\textbf{k}}|M_a\rangle = d_a(\hat{\textbf{k}})$. Often it is useful to represent the state $|\vec{\Delta}\rangle$ in the eigenstates of the third component of the orbital angular momentum, labeled $ |m\rangle $ with $m\in(-2, -1,0,1,2)$, such that $\langle \hat{\textbf{k}}|m\rangle = Y_{2m} (\theta,\phi)$ are the usual spherical harmonics.\cite{QM} The two representations are related through
\begin{align}
 |M_1\rangle &= \frac{1}{\sqrt{2}}\Bigl(|-2\rangle+|2\rangle\Bigr),\\
 |M_2\rangle &=  |0\rangle,\\
 |M_3\rangle &= \frac{1}{\sqrt{2}} \Bigl(|-1\rangle-|1\rangle\Bigr),\\
 |M_4\rangle &= \frac{\rmi}{\sqrt{2}} \Bigl(|-1\rangle+|1\rangle\Bigr),\\
 |M_5\rangle &= \frac{\rmi}{\sqrt{2}} \Bigl(|-2\rangle-|2\rangle\Bigr).
\end{align}
The basis states $|M_a \rangle$ are constructed to be invariant under time-reversal transformations and are in precise one-to-one correspondence with the five real Gell-Mann matrices $M_a$, which transform under SO(3), like the functions $d_a$, as components of a second-rank irreducible tensor\cite{boettcherSC,boettcher}. We have $d_a=(\sqrt{15}/2)(k_i M^a_{ij}k_j)/k^2$.

We factorize $\Delta_a$ into the overall norm $\Phi$  and the internal degrees of freedom $z_a$ through
\begin{equation}
\label{eq5} \Delta_a= \Phi^{1/2} z_a,
\end{equation}
with $z_a^* z_a  =1$. One can then interpret
\begin{align}
\label{eq5b} |\Psi\rangle = z_a |M_a\rangle
\end{align}
as the normalized internal quantum state of the Cooper pair and we have
\begin{align}
 \label{eq6} \langle \hat{\textbf{k}}|\Psi\rangle = d_a(\hat{\textbf{k}})z_a.
\end{align}
The definition of the $d_a$ functions then implies the normalization
\begin{equation}
\label{eq7} \langle \Psi|\Psi\rangle = \int \frac{\mbox{d}\Omega}{4\pi} \  |\langle \hat{\textbf{k}} | \Psi \rangle |^2 =1.
\end{equation}
 The average (orbital) magnetization of the state $|\Psi\rangle$ can be computed from the matrix $A=z_aM_a$ via
\begin{align}
 \langle \Psi|\textbf{L} |\Psi\rangle^2 = \sum_{i=1}^3 \langle \Psi| L_i|\Psi\rangle^2 = \frac{1}{2}\mbox{tr}\Bigl([A,A^\dagger]^2\Bigr).
\end{align}
The amplitude of the average magnetization is bounded from above by two in our units.

\section{Minimization at $T=0$}

\subsection{ Nonlinear eigenvalue problem}

We now rewrite the mean-field effective action in Eq. (\ref{eq3}) at $T=0$ as
\begin{equation}
\label{eq8} \frac{S[\vec{\Delta}]}{\cal{N}}= \frac{\Phi}{V} - \int '  \frac{\mbox{d}^2 Q}{2\pi} \int \frac{\mbox{d}\Omega}{4 \pi} \ln ( Q^2 + \Phi |\langle \hat{\textbf{k}} | \Psi \rangle |^2),
\end{equation}
where $\cal{N}$ is the density of states at the Fermi level, $Q= (\omega, \xi)$ with $\omega$ as the continuous frequency, and $V = g \cal{N}$ is the dimensionless coupling. Minimizing with respect to the norm $\Phi$ in the weak coupling regime $V\ll 1$ then yields the equation
\begin{equation}
 \label{eq9} F[\Phi_0] = \frac{1}{V} - \int \frac{\mbox{d}\Omega}{4 \pi} |\langle \hat{\textbf{k}}  | \Psi \rangle |^2 \ln \Bigl( \frac{ v_F \Lambda}{ \Phi_0 |\langle \hat{\textbf{k}}  | \Psi \rangle |^2}  \Bigr)  =0,
\end{equation}
with the solution
\begin{equation}
 \label{eq10} \Phi_0 =  v_F \Lambda e^{X - (1/V)}.
\end{equation}
Here $X$ is a functional of the normalized Cooper pair state given by
\begin{equation}
 \label{eq11} X[\Psi] = - \int \frac{\mbox{d}\Omega}{4 \pi}\ |\langle \hat{\textbf{k}}  | \Psi \rangle |^2 \ln |\langle \hat{\textbf{k}}  | \Psi \rangle |^2.
\end{equation}
It constitutes the central object of interest in this work. Subtracting the value in the normal phase, the difference in the action can be recast into
\begin{equation}
\label{eq12} \frac{S[\vec{\Delta}]}{\cal{N}} = \int_0 ^{\Phi_0[\Psi]} \mbox{d}\Phi\ F[\Phi].
\end{equation}
After the insertion of the solution for the norm $\Phi_0$ and some simple algebra we eventually arrive at
\begin{equation}
 \label{eq13} \frac{S_0[\Psi]}{\cal{N}}  = - \Phi_0[\Psi].
\end{equation}
The action in the weak-coupling regime is therefore a simple function of the normalized $\ell=2$ Cooper pair state through $p(\theta,\phi)=|\langle \hat{\textbf{k}} |\Psi \rangle|^2$ alone, and the ground state is evidently the one that {\it maximizes} the quantity $X$ and, together with it, the norm $\Phi_0$.

Optimizing $X$ under the normalization constraint in Eq. (\ref{eq7}) with the help of a Lagrange multiplier straightforwardly leads to the condition that any extremal solution $|\Psi\rangle$ of $X$ satisfies
\begin{equation}
\label{eq14} - \int \frac{\mbox{d}\Omega}{4\pi}  \ln |\langle \hat{\textbf{k}}  | \Psi \rangle |^2\ d_a (\hat{\textbf{k}})  d_b (\hat {\textbf{k}}) z_b    = X_0 z_a,
\end{equation}
where $X_0$ is the value of $X$ for this solution. This can be viewed as a nonlinear eigenvalue problem for the coefficients $z_a$. We are therefore after the solutions of the last equation, and in particular after the highest possible value of $X_0$. The absolute maximum of $X$ is reached for a rotationally invariant s-wave superconducting state, which corresponds to $X=0$. Consequently, $X<0$ for any $\ell=2$ state.

\subsection{Michel's theorem and the search for the ground state}

In search for the local extrema of the functional $X$ we consider Cooper pair states $|\Psi\rangle$ that are invariant under each allowed subgroup of SO(3) separately, and then maximize $X$ within each such class of states. Michel's theorem \cite{michel, bruder} then guarantees that each extremal state within these symmetric classes will automatically satisfy Eq. (\ref{eq14}).

With this principle in mind, consider first the smallest and hence the least restrictive discrete subgroup of SO(3), namely the $C_{2z}$ group of rotations by an angle of zero and $\pi$ around, for example, the z-axis. The two families of states that are eigenvectors of the nontrivial $C_{2z}$ transformation, with eigenvalues $-1$ and $+1$, are given by
\begin{equation}
\label{eq15} |\Psi_1 \rangle = c_+ |1 \rangle + c_- |-1\rangle
\end{equation}
and
\begin{equation}
\label{eq16} |\Psi_2 \rangle = a_+ |2\rangle + e^{\rmi\delta} b | 0\rangle + a_- |-2\rangle.
\end{equation}
The coefficients $c_\pm$, $a_\pm$, $b$ can be chosen to be real, since their phases could always be eliminated by a combined SO(3) and U(1) transformation. The remaining parameter $\delta$ can be taken in the range $0<\delta<\pi$. Normalization then leaves us with one and three real parameters to span the above two families of states, respectively.

For general values of the coefficients, the state $|\Psi_2\rangle$ has only $C_{2z}$ symmetry, but\cite{kawaguchi}: a) when $a_{\pm} =0$, it reduces to the uniaxial nematic state,\cite{boettcher} invariant under the continuous subgroup SO(2), b) when $b=a_-=0$ it becomes the ferromagnetic state with maximal average magnetization, also invariant under SO(2), c) when $b=0$, it is invariant under the subgroup $C_{4z}$, d) when $b=0$ and $a_+ = a_-$ it is invariant under the subgroup $D_4$, e) when $a_+ = a_-$ it is invariant under the subgroup $D_2$, and f) when $a_{\pm} = 1/2$, $b=1/\sqrt{2}$, and $\delta= \pi/2$ it becomes the {\it cyclic} state,\cite{mermin1} invariant under the tetrahedron group, $T_4$.

The above list leaves a single remaining subgroup of SO(3) under which an $\ell=2$  state can be invariant, but which is neither in the form of $|\Psi_1 \rangle$ or $|\Psi_2 \rangle$. This is $C_{3z}$, in which case the most general state modulo SO(3), U(1), and time reversal transformations can be written as \cite{kawaguchi}
\begin{equation}
\label{eq17} |\Psi_3 \rangle = d_+ |2 \rangle + d_- |-1\rangle
\end{equation}
with real coefficients $d_{\pm}$. This therefore defines the third and the final (one-parameter) family of states. One may note that for particular $d_+ = 1/\sqrt{3} $ and $d_- = \sqrt{ 2/3} $ the state $|\Psi_3 \rangle $  is in fact the same cyclic state as $|\Psi_2 \rangle$ in the case f), only nontrivially rotated.

Extremizing $X$ first within the one-parameter family of states $|\Psi_1 \rangle$ yields a minimum of $X_0 = -0.574717$ for $c_\pm= 1/\sqrt{2}$, i. e. for $|\Psi\rangle = -\rmi|M_4 \rangle$. In fact, one finds the same value of $X$ for any choice of $|\Psi\rangle = |M_a \rangle$, $a=1,3,4,5$, which are all real, biaxial nematic states, mutually related by SO(3) rotations. The maximum within the $|\Psi_1 \rangle$ family is $X_0= -0.267864$ for $c_- =0$, i. e. for $|\Psi\rangle = |1 \rangle$. Interestingly, the same value is obtained for any $L_3$ eigenstate $|m\neq 0 \rangle$, although the states with different values of $|m|$ are obviously not related by an SO(3) rotation.

Within the $|\Psi_3 \rangle $ family one finds that the maximum is $X_0= -0.222213$, which is the above-mentioned
cyclic state. The question is then whether this is the actual global maximum, or there are states within the remaining larger family of $|\Psi_2 \rangle$ which have a higher $X$. Somewhat surprisingly, the answer to the last question turns out to be positive, and we find the global maximum of $X$ to be reached for the particular state within the $|\Psi _2  \rangle$ family given by
\begin{equation}
\label{eq18} |\Psi_{\rm opt}  \rangle = 0.898816 |2\rangle + \rmi 0.432951 |0\rangle + 0.068431 |-2\rangle,
\end{equation}
which satisfies the nonlinear eigenvalue Eq. (\ref{eq14}), and has the highest value of $X_0 = -0.206173$ among all such solutions.

Besides having only the minimal residual $C_2$ symmetry this state altogether appears quite unexceptional. It breaks time reversal symmetry, but not maximally, since the overlap between the state and its time-reversed copy is
\begin{equation}
\label{eq19} \langle \Psi_{\rm opt} |\hat{T} |\Psi_{\rm opt} \rangle = -0.0644
\end{equation}
and, although small, not quite zero. Its average orbital magnetization is
\begin{equation}
\label{eq20} |\langle \Psi_{\rm opt} |\textbf{L}|\Psi_{\rm opt} \rangle |  = 1.60637,
\end{equation}
and, although high, below the maximal value of two.

\section{Ginzburg-Landau theory at $T\neq 0$ }

\subsection{ Sixth-order expansion}

At $T\neq 0$ the Ginzburg--Landau expansion of the action becomes necessary, and it is instructive to compare the ground state obtained in this manner with the exact result. Expanding the logarithm in Eq. (\ref{eq3}) in powers of $\Delta_a$ gives
\begin{align}
 \nonumber S[\vec{\Delta}] ={}&  r |\vec{\Delta}|^2 + q_1  |\vec{\Delta}|^4 + q_2 |\vec{\Delta}^2|^2 +
  s_1 |\vec{\Delta}|^6 \\
\label{eq21} &+ s_2 |\vec{\Delta}|^2 | \vec{\Delta}^2|^2 + s_3 Y (y) + O(\Delta^8),
\end{align}
where the coefficients $r$, $q_1$ and $q_2$ can be discerned easily. One can check that $q_1 = 2 q_2 >0$, and the quartic term favors the configurations with
\begin{equation}
\label{eq22}  \vec{\Delta}^2=\Phi \langle \Psi |\hat{T} |\Psi \rangle=0,
\end{equation}
i. e. with {\it maximal} breaking of time reversal symmetry.\cite{boettcher} The third symmetry-allowed quartic term,  \cite{boettcher, mermin}
which in matrix notation $A=z_aM_a$ introduced earlier would be proportional to
$\mbox{tr}(A^\dagger A A^\dagger A) $ is, however, absent, and it is left to the sextic terms to remove the degeneracy between the maximally time-reversal-symmetry-breaking solutions of Eq. (\ref{eq22}). The only  sextic term capable of doing so is the last one in Eq. (\ref{eq21}), which reads
\begin{equation}
\label{eq23} Y(y) = | \mbox{tr} A^3 | ^2 + y  | \mbox{tr}(A^2A^\dagger) | ^2.
\end{equation}
We find that $s_2=3s_1 /2<0$, $s_3 <0$, and the {\it relative} coefficient between the two terms in $Y$ is $y=9$. This particular value results as a property of the integral over the products of six functions $d_a (\hat{\textbf{k}})$, in a similar way to the relative coefficient of $2$ between $q_1$ and $q_2$ in the quartic term. Its large numerical value, however, turns out to be crucial in determining the ground state, as we explain next.

For general $y$ the configuration with $\vec{\Delta}^2 =0$ that  maximizes $Y(y)$ may be cast into the (normalized) form
\begin{equation}
\label{eq24} \vec{\Delta} = \frac{1}{\sqrt{2} } ( 1,\rmi \sin \alpha, 0,0, \rmi \cos\alpha),
\end{equation}
for which the pertinent sextic term $Y(y) $ becomes
\begin{equation}
\label{eq25} Y(y) = \frac{8}{3} (\sin \alpha)^2 [ (\sin\alpha)^4+ y (\cos \alpha)^4] .
\end{equation}
For $y<6.46$ the maximum of $Y(y)$ is at $\alpha=\pi/2$, which is the cyclic state, whereas for $y>6.46$ it shifts to the state with
\begin{equation}
\label{eq26} (\sin\alpha )^2 = \frac{2y-\sqrt{ y^2-3y}}{3(y+1)}\approx \frac{1}{3} \Bigl[ 1+ \frac{1}{2y} + O\Bigl(\frac{1}{y^2}\Bigr) \Bigr].
\end{equation}
Since the actual value is $y=9$, it is the latter state that wins over the cyclic state. To compare it with the exact ground state let us write the approximate Ginzburg--Landau state in Eq. (\ref{eq24}) in the angular momentum basis as
\begin{equation}
\label{eq27} |\text{GL} \rangle = \frac{1+\cos\alpha}{2} |2\rangle + \frac{\rmi\sin \alpha }{\sqrt{2}} |0\rangle + \frac{1-\cos\alpha}{2} |-2\rangle.
\end{equation}
Taking then the solution of Eq. (\ref{eq26}) with  $\sin\alpha= \sqrt{(6-\sqrt{6})/10}$ and $\cos\alpha = \sqrt{1- (\sin\alpha)^2}$ yields $X=-0.207261$ and a large overlap with the exact state:
\begin{equation}
\label{eq28} \langle \Psi_{\rm opt}  | \text{GL}\rangle = 0.99948.
\end{equation}
The average magnetization of the Ginzburg--Landau state is similarly close:
\begin{equation}
\label{eq29} |\langle \text{GL} |\textbf{L}|\text{GL} \rangle|  = 2 \cos\alpha= 1.60617.
\end{equation}
Even taking only the first term in the expansion for large $y$ in Eq. (\ref{eq26}) yields already a similar agreement.  Taken together, the Ginzburg--Landau state is an excellent approximation to the exact ground state, and therefore in the weak-coupling limit the system is essentially in one and the same superconducting state at {\it all} temperatures below the critical temperature. One can also show that the last state in the Table 1 in Ref. \cite{boettcher} is the same as $|\text{GL}\rangle$, modulo U(1) and SO(3) transformations.

We have also conducted a random sampling of the entire $\ell=2$ Hilbert space in search of a higher value of $X$ and found none. As an additional check of our procedure, one can also compute the value of $X$ over the $\ell=1$ states. Since factoring out U(1) and SO(3) in that case leaves only one real parameter, it is easy to locate all the extremal values without the use of Michel's theorem. We find the single maximum of $X= -0.125$ in the ferromagnetic state $|1\rangle$ and the minimum of $X= -0.431$ in the uniaxial nematic state $|0\rangle$. Both have a residual SO(2) symmetry  and, therefore, the use of Michel's theorem in this case would find the same result.

\subsection{ Coupling to parasitic s-wave}

Finally, we wish to point out that the coupling to the  s-wave component,\cite{kim} although allowed, does not really change the ground state. For this include the terms that contain the s-wave superconducting order parameter $\Delta_0$ in the Ginzburg-Landau expansion, so that the expanded action to the two lowest  orders is changed to
\begin{align}
\label{eq30} S'= S + r_{\rm s} |\Delta_0|^2 + u \Bigl( \Delta_0 ^* \mbox{tr}(A^2A^\dagger) + \text{c.c} \Bigr).
\end{align}
The quadratic coefficient is assumed positive ($r_{\rm s} >0$), so that there is no s-wave order when $\Delta_a =0$. The coefficient $u$ vanishes by particle-hole symmetry in the four-band Luttinger problem,\cite{boettcher}  but it is finite away from the particle-hole symmetric point, and should be naturally included when $\mu\neq 0$. \cite{kim} Since $r_{\rm s} >0$, one may simply perform the Gaussian integration over  the s-wave component, with the main effect being a change in the sextic term for the  remaining d-wave component given by
\begin{equation}
\label{eq31} Y(y) \rightarrow Y\Bigl(y+ \frac{u^2}{r_{\rm s}}\Bigr).
\end{equation}
The already large value of $y$ is only {\it increased} by the coupling to the s-wave component, which therefore does not alter the ground state, except in a negligible quantitative way.

\section{Discussion}

In conclusion, we have determined the exact mean-field ground state of the BCS d-wave superconductor at $T=0$, and shown that it is quite robust to the effects of finite temperature and mixing with an s-wave component. The state has the same symmetry and is even quantitatively close to the one obtained from the sixth order Ginzburg--Landau expansion that was proposed in Refs. \cite{boettcher} and \cite{kim}. It breaks time-reversal symmetry, reduces the rotational symmetry down to the $C_{2}$, and has a large magnetization. The quasiparticle spectrum (Eq. (6)) consists of point nodes, and has been studied in \cite{kim}.

Note that the ground state $|\Psi_{opt} \rangle $ is the only {\it noninert} state among all the extremal points of the energy functional we discussed. All other extremal states, namely the biaxial nematic $| M_{m\neq2} \rangle$ ($D_4$-symmetric),  uniaxial nematic $| M_2 \rangle$ ($SO(3)$-symmetric ), ferromagnetic states $|m=2\rangle$ ($SO(3)$-symmetric), and $|m=1\rangle$ ($SO(3)$-symmetric), and the cyclic state ($T_4$-symmetric) are unique states with their respective symmetries\cite{remark}, i. e. they are ``inert states". As such they are, by the Michel's theorem, the saddle points of {\it any} $SO(3)$-symmetric functional. $|\Psi_{opt}\rangle $, however, is obviously only one among many states with the $C_2$ symmetry, and had to be found by maximization with respect to three real parameters. It is therefore nongeneric, and tied to the BCS form of the energy, i. e. to the specific form of the functional $X[\Psi]$.

Comparing the maximal value of $X$ in the ground state $(-0.206173)$ with the next highest local maximum $(-0.222213)$ attained in the cyclic state, we see that they differ by only a few percent. This means that the effects of thermal and quantum fluctuations that lie beyond the mean-field theory considered here may be significant in determining the ground state configuration.\cite{rg} This will be the topic of a separate publication. \cite{herbut}

\section{Acknowledgments}

 This work was supported by the NSERC of Canada, the DoE BES QIS program (award No. DE-SC0019449), the NSF PFCQC program, AFOSR, DoE ASCR Quantum Testbed Pathfinder program (award No. DE-SC0019040), ARO MURI, ARL CDQI, and NSF PFC at JQI.

\end{document}